\documentclass[entropy,article,accept,moreauthors,pdftex]{Definitions/mdpi}

\newcommand{\hx}[2]{\ensuremath{h_{\times}(#1 \mid #2)}}
\def\ER{Erd{\H o}s-R{\'e}nyi }
\def\BA{Barab{\'a}si-Albert }

\firstpage{1} 
\pubvolume{22}
\issuenum{3}
\articlenumber{265}
\pubyear{2020}
\copyrightyear{2020}
\history{Received: 1 February 2020; Accepted: 24 February 2020; Published: 26 February 2020}
\Title{Complex Contagion Features without Social Reinforcement in a Model of Social Information Flow}

\Author{Tyson Pond $^{1}$, 
Saranzaya Magsarjav $^{2}$,
Tobin South $^{2}$,
Lewis Mitchell $^{2}$ and 
James P.\ Bagrow $^{1,}$*}

\AuthorNames{Tyson Pond, Saranzaya Magsarjav, Tobin South, Lewis Mitchell and James P. Bagrow}

\address{%
$^{1}$ \quad Department of Mathematics \& Statistics, University of Vermont, Burlington, VT 05405, USA\\
$^{2}$ \quad School of Mathematical Sciences, The University of Adelaide, Adelaide SA 5005, Australia}

\corres{Correspondence: james.bagrow@uvm.edu }

\abstract{Contagion models are a primary lens through which we understand the spread of information over social networks.
However, simple contagion models cannot reproduce the complex features observed in real-world data, leading to research on more complicated complex contagion models.
A noted feature of complex contagion is social reinforcement that individuals require multiple exposures to information before they begin to spread it themselves.
Here we show that the quoter model, a model of the social flow of written information over a network, displays features of complex contagion, including the weakness of long ties and that increased density inhibits rather than promotes information flow.
Interestingly, the quoter model exhibits these features despite having no explicit social reinforcement mechanism, unlike complex contagion models.
Our results highlight the need to complement contagion models with an information-theoretic view of information spreading to better understand how network properties affect information flow and what are the most necessary ingredients when modeling social behavior.
}

\keyword{online social networks;
social media;
information spreading;
information diffusion;
cross-entropy
}

\begin{document}

\section{Introduction}

Social networks mediated through online platforms are an increasingly important way in which individuals send and receive information, and their influence is now felt in economics, politics, and the workplace~\cite{lazer2009computational,tumasjan2010predicting,conover2013digital,castells2015networks,de2015unique,garcia2017leaking}.
These platforms provide rich opportunities for researchers to collect and study real-world data related to human behavior and the spread of information.
In concert with these datasets, considerable research has worked towards better statistical and information-theoretic tools to quantify information flow~\cite{schreiber2000measuring,sun2014causation,borge2016dynamics} and towards more accurate mathematical models to understand and even predict information flow~\cite{wang2011information,bagrow2019information,quoterModel2018}.

A common approach to measuring information flow over a network is to idealize information as a collection of `packets,' and then track the spread of those packets throughout the network. 
This approach is especially common when studying social media where keywords such as hashtags or URLs are easily tracked.
More complex phenomena, such as the adoption of behaviors can also be monitored and used as a proxy for information flow~\cite{centola2010spread}.
Treating information flow in this way brings to mind the spread of infections and the use of epidemiologically inspired models is popular.
In this context, the social ``diffusion'' of information is often characterized as either a simple contagion or a complex contagion~\cite{10.1093/comnet/cnt006}.
Simple contagions are those where each exposure can independently lead to an infection.
Complex contagions, in contrast, introduce a social reinforcement mechanism where multiple exposures are needed before the contagion can spread.

However, despite its simplicity and popularity, there can be drawbacks to treating information as the contagion of discrete packets. 
Within social media, for example, there is a wealth of written information being posted by users that is ignored when focusing only on particular keywords.
Likewise, considerable information could be exchanged between individuals without leading to an observable adoption of behavior.
Therefore, we argue in this work that a more nuanced approach grounded in information theory can give a better view of information flow in online social networks while more fully using the available data.

The goal of this work is to study how network properties can affect information flow when taking an information-theoretic view on information flow, and how this information-theoretic view compares to contagion.
We study the quoter model~\cite{quoterModel2018}, a simple model for individuals generating text data within social media and apply information-theoretic estimators to the model text.
Using both network models and real-world network data, we compare the behavior of information flow in this model with traditional simple and complex contagion, to see the similarities and differences we may observe through these contrasting viewpoints.
Interestingly, we find that the quoter model exhibits several phenomena characteristic of complex contagion, despite lacking an explicit social reinforcement mechanism, the key feature of complex contagion.

The rest of this work is organized as follows.
In Section~\ref{sec:background} we describe information-theoretic estimators of information flow and mathematical models of information flow and contagion.
In Section~\ref{sec:methods} we describe the materials and methods used in this study, including simulation details, measures of information flow, the network properties we investigate, and the network data we use.
Section~\ref{sec:results} presents our results comparing contagion models with the information-theoretically motivated quoter model and exploring how various network properties affect information flow in the quoter model. 
We conclude with a discussion in Section~\ref{sec:discussion}.

\section{Background}
\label{sec:background}

\subsection{Measuring Information Flow}
\label{subsec:measuringinfoflow}

Suppose an individual within a social network generates a stream of text representing posts shared online on Twitter, for example.
The entropy rate $h$ of this text captures the information present within it.
It can be challenging to estimate $h$ for natural language data as information is present in the ordering of the words, not just the relative frequencies of words~\cite{shannon1951prediction}.
To help address this challenge,
Kontoyianni et al.~\cite{kontoyiannis1998nonparametric} proved that the estimator
\begin{equation}
    \hat{h} = \frac{T \log_2 T}{\sum_{t=1}^{T} \Lambda_{t}}, %
    \label{eqn:hhat}
\end{equation}
converges to the true entropy rate $h$ of a text,
where $T$ is the length of the sequence of words and $\Lambda_{t}$ is the \emph{match length} of the prefix at position $t$: it is the length of the shortest substring (of words) starting at $t$ that has not previously appeared in the text. 
This estimator has been used to study human dynamics including mobility patterns and social media predictability~\cite{song2010limits,bagrow2019information}.

Equation~\eqref{eqn:hhat} generalizes to an estimator of the \textbf{cross-entropy} $h_\times$ between two texts $A$ and $B$~\cite{ziv1993measure,bagrow2019information}:
\begin{equation}
	\hat{h}_{\times}(A \mid B) = \frac{T_{A} \log_2 T_{B}}{\sum_{t=1}^{T_{A}} \Lambda_{t}(A \mid B)},
	\label{eqn:crossEntropy}
\end{equation}
where $T_{A}$ and $T_{B}$ are the lengths of the two texts, and $\Lambda_{t}(A | B)$ is the length of the shortest substring $[A_t, A_{t+1}, \ldots, \allowbreak A_{t+\Lambda_t(A\mid B)+1}]$ starting at position $t$ of text $A$ not previously seen in text $B$. 
Previously, in this case, refers to all the words of $B$ written prior to the time when the $t$th word of $A$ was written.
Specifically, compute $\Lambda_{t}(A | B)$ by searching for each  substring $[A_t]$, $[A_t, A_{t+1}]$, $...$ within $B_{:t} \equiv [B_j \mid \mathrm{time}(B_j) < \mathrm{time}(A_t)]$, the ordered sequence of words in $B$ that appear before the time of the $t$-th word in $A$, until the first substring $[A_t, \ldots, A_{t+\Lambda_t(A\mid B)+1}]$ that is not seen in $B_{:t}$.
By matching the future text of $A$ (words posted at times $\geq \mathrm{time}(A_t)$) against the past text of $B$ (words posted at times $< \mathrm{time}(A_t)$) at every $t$, only $B$'s past predictive information about $A$'s future is estimated and \emph{temporal precedence} is satisfied.
The cross-entropy can be applied directly to the texts of a pair of individuals by choosing $B$ to be the text stream of one individual and $A$ the text stream of the other, and Equation~\eqref{eqn:crossEntropy} can be used to measure the information flow between those individuals by asking how much predictive information about one text is contained within the other.
This can be a quite powerful and effective measure of information flow, as it satisfies temporal precedence of the text streams and it uses all of the available (text) data for the pair of users~\cite{kontoyiannis1998nonparametric,ziv1993measure,schreiber2000measuring,quoterModel2018,bagrow2019information}.

We focus on the cross-entropy estimated using Equation~\eqref{eqn:crossEntropy} as a pairwise measure of information flow, but generalizations can capture information flow from multiple social ties towards a single individual~\cite{quoterModel2018,bagrow2019information}.
Doing so allows for measures of more complex information flow such as analogs of transfer entropy or causation entropy~\cite{schreiber2000measuring,sun2014causation,sun2015causal}.
The best extensions of information flow estimators beyond pairwise measures remains an active and fruitful area of research (see also our discussion in Section~\ref{sec:discussion}).

Closely associated with the cross-entropy is the predictability $\Pi$.
Predictability, given by Fano's Inequality~\cite{cover2012elements}, provides a bound on how accurately an \emph{ideal} predictive method can perform when working with data of a given entropy: $\Pi$ is the probability the most accurate possible method will correctly predict the subsequent word with the given information's uncertainty (i.e., the cross-entropy).
\begin{equation}
    h(\Pi)+(1-\Pi) \log (z-1) \geq h_\times
    \label{eqn:fanos}
\end{equation}
where $h(\Pi) = -\Pi \log (\Pi)-(1-\Pi) \log (1-\Pi)$ and $z$ is the cardinality of the sample space; in our problem, this is the vocabulary size or number of unique words for the quoter model (Section~\ref{subsec:simulatingquotermodel}).
The predictability is then given by finding numerically the largest $\Pi$ that satisfies Equation~\eqref{eqn:fanos}.
Equation \eqref{eqn:fanos} demonstrates that $h_\times$ and $\Pi$ are functionally equivalent (and inversely related, with higher $h_\times$ corresponding to lower $\Pi$ and vice versa) as $z$ is a constant for the model we study here (see also discussion in Section~\ref{sec:discussion}).
Higher values of $\Pi$ (lower $h_\times$) correspond to higher amounts of information flow.

\subsection{Quoter Model}
\label{subsec:quoter}

To study the effects of network properties on information flow, we use the recently proposed quoter model~\cite{quoterModel2018}.
The quoter model represents an idealized model of social conversations, meant to capture some of the processes by which individuals in an online social network post text while also being analytically tractable.
Nodes in a network generate text streams both by sampling from a given vocabulary distribution and by copying (``quoting'') short sub-sequences of text from their neighbors. 
This model provides a parameter $q$, the quote probability that tunes the degree of information flow.
(Full details of the model and how we simulate it are given in Section~\ref{subsec:simulatingquotermodel}.)
After simulating the quoter model for a given number of time steps (Section~\ref{subsec:simulatingquotermodel}), a text stream has been generated by each node in the network, and we can estimate the cross-entropies between these texts to study the social flow of written information.
See Bagrow and Mitchell~\cite{quoterModel2018} for full details on the quoter model.

\subsection{Other Models of Information Flow}
\label{subsec:existingmodelsinfoflow}

Contagion approaches are often used to model information flow~\cite{10.1093/comnet/cnt006}.
A classic simple contagion approach to information flow is compartment models, taken from models of epidemics. 
Two simple compartment models are Susceptible-Infected (SI) and Susceptible-Infected-Recovered (SIR) models.
On a network, a small number of nodes are initially ``infected'' while the remaining nodes are susceptible. The contagion then spreads from those infected nodes with a constant transmission rate per link so that each node in the ``S'' compartment has a constant probability to move to the ``I'' compartment with any given exposure.
For SIR models, an additional ``R'' compartment is used to model a recovery process where infected nodes cease spreading the contagion while also becoming immune to reinfection. 
Many variants on these models exist.

Complex contagion phenomena are typically captured with threshold models~\cite{granovetter1978threshold,watts2002simple}.
Here nodes are again labeled as susceptible or infected, but the probability for a node $i$ to become ``infected'' is a function of the number of neighbors of that node already infected.
If too few neighbors are infected there is zero probability that $i$ will be infected.
Yet if a sufficient fraction of $i$'s neighbors become infected, then $i$ has a non-zero probability of becoming infected.
This \emph{social reinforcement} mechanism is intended to capture the cognitive mechanisms underlying opinion change, knowledge acquisition, and other facets of how individuals respond to and adopt information and ideas
\cite{centola2007cascade,Ugander5962}.

Complex contagion leads to several phenomena that differ from simple contagion.
For one, there is an interesting \emph{cascade window} where network density leads to a non-monotonic relationship with the spread of the contagion.
Often denser networks lead to less spread, unlike simple contagion where a contagion will spread more easily as denser networks afford more opportunities (links) for spreading.
Another feature of complex contagion is the complicated role of clustering
where clustering can appear to either promote or inhibit contagion~\cite{miller2009randomClustered,pastor2015epidemic,o2015mathematical,gray2018super}.
Complex contagion also exhibits a ``weakness of long ties'' effect, where long ties impede the flow of contagion~\cite{centola2007complex}, in contrast with the seminal ``strength of weak ties'' result~\cite{granovetter1977strength} that implies long-range ties have an out-sized role in promoting information flow.
The goal of our work here is to study the information-theoretic view of information flow we adopt here with the quoter model and compare to the effects of complex contagion that is commonly used as a \emph{non}-information-theoretic view to study information flow.

\section{Materials and Methods}
\label{sec:methods}

In this study, we use the quoter model on networks to elucidate the role of network structure on information flow. 
Here we describe the procedures to simulate the quoter model, measure information flow between nodes in networks, we describe the network features we study in relation to information flow, and we provide the details on the network models (random graphs) and real-world network datasets we study.

\subsection{The Quoter Model}
\label{subsec:simulatingquotermodel}
We use the following process to simulate the quoter model on a given network.
The quoter model requires a directed graph $G=(V,E)$ (where $N = \left|V\right|$ is the number of nodes and $M=\left|E\right|$ is the number of edges) and, in the most general case, quote probabilities $q_{uv}$ on each directed edge (we say node $v$ (ego) may quote $u$ (alter) if the edge $u \to v$ exists and has $q_{uv} > 0$).
We simplify this for our simulations: when an ego generates new text, with probability $q$ (bidirectional quoting) we pick an alter (predecessor) uniformly at random to quote from; otherwise, with probability $1-q$ the ego generates new content. 
If an ego quotes an alter (probability $q$), copy a random segment of the alter's past text and append this onto the ego's growing text stream. 
We take the ``quote length'' (number of words) being copied to be Poisson-distributed (with mean $\lambda$) for all users; %
Otherwise, if not quoting (probability $1-q$), generate new content by sampling with replacement from a vocabulary distribution $W(w)$ and appending those samples onto the ego's growing text stream, where the number of samples is again Poisson-distributed with mean $\lambda$.
We assume a common, fixed vocabulary distribution $W(w)$ that follows a Zipf law of word use, as in prior studies and motivated by real-world language usage patterns~\cite{quoterModel2018}. 
Specifically, a Zipf law defines the probability of using word $w$ to be a power law based on the rank $r_w$ of $w$:
$W(w) = H_{z,\alpha}^{-1} r_w^{-\alpha}$, 
where $z$ is the vocabulary size and $H_{z,\alpha} = \sum_{r=1}^z r^{-\alpha}$.
Here we take $z=1000$ as in \cite{quoterModel2018} and, unless otherwise stated,
focus on the exponent $\alpha = 1.5$, a value typical of social media data.
We focus in this work on $q=1/2$ and $\lambda=3$ but we explore the robustness of our results to other parameter choices in Appendix~\ref{app:robustness}.
This process repeats for $T=1000N$ time steps so that each user has generated approximately $1000\lambda = 3000$ words when complete.
This number of time steps was chosen to ensure the entropy estimator would converge (see~\cite{kontoyiannis1998nonparametric,ziv1993measure} for convergence proofs).
While very short amounts of text will make the estimated entropy too uncertain to be reliable,
this length of text is in line with the empirical  convergence of $h_\times$ reported in real data~\cite{bagrow2019information}.

\subsection{Measuring Information Flow over the Network}
\label{subsec:measuringinfoflowMethods}

After generating text streams for all nodes in $G$ by iterating the quoter model, the cross-entropy estimator  (Equation~(\ref{eqn:crossEntropy})) is then applied as needed to compute $h_\times$.
We compute the cross-entropy over all edges, $\{h_\times\} = \{ \hx{u}{v} \mid (u,v) \in E\}$, and report the mean $\left<h_\times\right>$ and variance $\mathrm{Var}(h_\times)$ of these values.
(We examine the distribution of $h_\times$ in Appendix~\ref{app:distr} to show that $\left<h_\times\right>$ and $\mathrm{Var}(h_\times)$ are reasonable summaries of the distribution of $h_\times$.)
Likewise, the predictability $\Pi$, given by Fano's Inequality~\cite{cover2012elements}, is a functionally equivalent measure of information flow (as we assume the same vocabulary sizes for nodes in the quoter model).
We focus on link-based cross-entropies although the cross-entropy estimator can be applied to non-neighboring nodes. 
Indeed, when studying the role of community structure in modular networks (see Section~\ref{subsec:syntheticnetworks}), we also consider cross-entropies between nodes in different modules, to assess information flow between and within said modules.

\subsection{Simulating Contagion Models}

To compare and contrast information flow in the quoter model, we also simulate traditional models of information flow, specifically simple and complex contagion.
For simple contagion we simulate a stochastic SIR model on different networks (1000-node Erd{\H o}s-R{\'e}nyi and Barab{\'a}si-Albert networks, as well as a sample of real-world networks) using \cite{miller2019eon}.
For the simulations here we set the transmission rate 20 and recovery rate 1.
We initialize with a random 5\% of the nodes infected, and run 10 outbreaks on 100 realizations of the network for each choice of average degree $\langle k \rangle$.
For complex contagion we use exactly the same parameters, except we introduce a threshold function for transmission as in \cite{watts2002simple}, where the transmission rate is set to zero if the proportion of infected neighbors is below some threshold $\phi$ (and we set $\phi = 0.18$ following \cite{watts2002simple}).
For all simple and complex contagion simulations we measure the peak outbreak size,
noting that larger outbreak sizes conventionally correspond to greater information flow.

\subsection{Assessing the Impact of Structure on Dynamics}
\label{subsec:syntheticnetworks}

In this work we use several network models (random graphs) tailored to control for various network properties such as density, clustering, and modular structure.
Here we describe the models and properties we study in relation to information flow in the quoter model.
\paragraph{Density and Average Degree}
To explore how network density relates to information flow, we create \ER and \BA networks of $N$ nodes with varying average degree, $\langle k \rangle$, allowing us to the tune their densities.
For the \ER networks we add edges independently with probability $p=\langle k \rangle /(N-1)$. 
For the \BA model we start with $m=\langle k \rangle /2$ nodes with no edges and add nodes which each form $m$ links with previous nodes according to preferential attachment.
Here we measure how cross-entropies varies with the densities of the networks using their average degree $\left<k\right>$ and edge density $M / \binom{N}{2}$ where $M$ is the total number of edges in the network.
To complement the \ER and \BA results, we also compare the densities of real networks with their average cross-entropy.

\paragraph{Degree Heterogeneity}
To assess the role of degree heterogeneity on information flow, we study the simplest random graph model with tunable degree heterogeneity, termed ``dichotomous networks'' in \cite{lambiotte2007does}. 
Dichotomous networks are generated via the configuration model. 
They have only two types of nodes -- those with degree $k_1$ and those with degree $k_2$. 
We assume there are $N/2$ nodes of each degree and fix $k_1+k_2$ so that the average degree is fixed. 
The mean and variance of the degree distribution, respectively, are given by $\mu = \frac{1}{2}\left(k_1+k_2\right)$ and $\sigma^2 = (k_1-k_2)^2/4$.
We are interested in how the cross-entropy varies with $k_1/k_2$. 
When $k_1/k_2=1$ the network reduces to a random $k$-regular graph ($\sigma^2$ = 0), while $\sigma^2 \to \infty$ as $k_1/k_2 \to 0$.

\paragraph{Clustering}
Clustering or triadic closure, the tendency towards forming triangles, is a key feature of social networks. 
We studied clustering using a network model with tunable numbers of triangles and with a randomization procedure that can lower the number of triangles in an existing network.
We quantify a network's clustering using \emph{transitivity} $T(G)$, the fraction of possible triangles in the network which actually exist: $T(G) = 3 N_\mathrm{triangles} / N_\mathrm{triads}$, where $N_\mathrm{triangles}$ counts the number of triangles in the network and $N_\mathrm{triads}$ is the number of triads or paths of length 2.

We constructed ``small-world'' networks using the Watts–Strogatz (WS) model \cite{watts1998collective} to tune their clustering. 
We generated a one-dimensional periodic lattice of $N$ nodes with $k$ nearest-neighbor connections, and randomly rewired lattice edges with a rewiring probability $p$.
Varying the rewiring probability $p$ allows us to tune the network diameter and clustering. 

While the Watts–Strogatz model lets us generate networks with different clustering values,
a generic challenge when assessing the impact of clustering (and other network properties) on dynamics is generating networks with tunable clustering, but for which other structural properties, such as density or diameter, can be controlled for. 
To study the relationship between transitivity and information flow, we apply the established degree-preserving stochastic rewiring or ``x-swap'' method~\cite{Singh:2013aa,milo2003uniform,diaconisXswaps}, in which we repeatedly choose two links at random and two randomly selected endpoints of those links are swapped as long as the number of links does not change by swapping and the network does not become disconnected. 
These swaps lower transitivity while fixing the number of links and degrees of all nodes in the network.
We performed $5M$ swaps for each real network.
Examining information flow on the randomized network compared with information flow on the original network can then illustrate what effect, if any, transitivity had on information flow.
\paragraph{Community Structure and Modularity}
Community structure is another inherent property of social networks. It is commonly quantified using modularity~\cite{newman2004finding}:
$$Q = \frac{1}{2M}\sum_{i,j}\left(a_{ij} -\frac{k_i k_j}{2M}\right)\delta(c_i,c_j), $$
where 
$M$ is the total number of links, 
the sum runs over all pairs of nodes in the network,
$\mathbf{A}=[a_{ij}]$ is the adjacency matrix of the network, 
$k_i$ is the degree of node $i$, 
$\delta$ is the Kronecker delta,
and
$c_i$ denotes the community containing $i$.
The community structure encoded in the $\{c_i\}$ can be found using a community detection algorithm or it may be planted within a network model.
To investigate community structure within a network model, we examined instances of the stochastic block model (SBM)~\cite{Danon_2005,PhysRevE.83.016107} with $N$ nodes and two planted blocks, or groups of nodes, denoted $A$ and $B$, of equal size $m=N/2$. 
Here there are two connection probabilities: $p_0$ (the within-block connection probability) and $p_1$ (the between-block connection probability) governing the probability for a link to form between nodes in the same block and in different blocks, respectively.
The expected modularity in this two-block stochastic block model is
$$Q = \frac{1}{2}\left(\frac{p_0 - p_0 m + p_1 m}{p_0 - p_0 m - p_1 m} \right).$$
Our main quantities of interest are the average cross-entropy on within-block edges, $\langle h_\times(\text{within}) \rangle$, the average cross-entropy on between-block edges $\langle h_\times(\text{between}) \rangle$ and their difference, $\Delta h_\times \equiv \langle h_\times(\text{between})\rangle - \langle h_\times(\text{within}) \rangle$. 
These quantities describe to what extent information flows within and between communities.

We also computed modularity for real networks using the Louvain method~\cite{blondel2008fast}. 
The Louvain method is a hierarchical community detection algorithm that finds a partition of nodes that maximizes modularity $Q$. 
As commonly done, we initialize each node in its own community.

\paragraph{Multiple Vocabulary Distributions}
A recent study \cite{deArruda2020impact} showed that heterogeneity in the dynamical parameters can be as important as structural heterogeneity.
Communities offer an obvious way to implement such heterogeneity:
We also investigate a two-block SBM where we distinguish the two groups $A$ and $B$ by giving them different Zipf exponents $\alpha_A,\alpha_B$, respectively, for their vocabulary distributions.

\subsection{Network Datasets}
\label{subsec:datasets}

To supplement the above graph models, we also studied contagion and quoter model dynamics on real-world networks. 
We developed a corpus of 10 social networks spanning a range of sizes and densities that were used as the basis for simulation.
See Appendix~\ref{app:netcorpus} for details on network sources and processing. 
Table \ref{tab:real_networks} shows several descriptive statistics for the networks we analyzed.

\begin{table} [H]
\centering
\caption{Descriptive statistics for real-world networks used in this study.
ASPL: Average Shortest Path Length.
Modularity computed using the Louvain method~\cite{blondel2008fast}.
\label{tab:real_networks}}
\begin{tabular}{ccccccccc}
	\toprule[1.5pt]
	{\bf Network} & \boldmath$|V|$ & \boldmath$|E|$ & \boldmath$\langle k \rangle$ & {\bf Density} & {\bf Transitivity} & {\bf ASPL} & {\bf Modularity} & {\bf Assortativity}\\
	\hline
    Sampson's monastery & 18 & 71 & 7.9 & 0.464 & 0.53 & 1.54 & 0.29 & $-$0.07\\
    Freeman's EIES & 34 & 415 & 24.4 & 0.740 & 0.82 & 1.26 & 0.07 & $-$0.15\\
    Kapferer tailor & 39 & 158 & 8.1 & 0.213 & 0.39 & 2.04 & 0.32 & $-$0.18\\
    Hollywood music & 39 & 219 & 11.2 & 0.296 & 0.56 & 1.86 & 0.20 & $-$0.08\\
    Golden Age & 55 & 564 & 20.5 & 0.380 & 0.53 & 1.64 & 0.45 & $-$0.13\\
    Dolphins & 62 & 159 & 5.1 & 0.084 & 0.31 & 3.36 & 0.52 & $-$0.04\\
    Terrorist & 62 & 152 & 4.9 & 0.080 & 0.36 & 2.95 & 0.52 & $-$0.08\\
    Les Miserables & 77 & 254 & 6.6 & 0.087 & 0.50 & 2.64 & 0.56 & $-$0.17\\
    CKM physicians & 110 & 193 & 3.5 & 0.032 & 0.16 & 4.24 & 0.61 & \phantom{$-$}0.11\\
    Email Spain & 1133 & 5452 & 9.6 & 0.009 & 0.17 & 3.61 & 0.57 & \phantom{$-$}0.08\\
	\bottomrule[1.5pt]
\end{tabular}
\end{table}

\section{Results}
\label{sec:results}

Here we compare information flow in the quoter model with traditional simple and complex contagion (Section~\ref{subsec:density}), then investigate how degree heterogeneity (Section~\ref{subsec:density}), clustering (Section~\ref{subsec:clustering}) and network modularity (Section~\ref{subsec:modularity}) affect information flow.
We also study how heterogeneity in the parameters affects information flow compared to the effects of network structure (Section~\ref{subsec:dynamicheterogeneity}).

\subsection{Information Flow and Models of Contagion}
\label{subsec:density}

\begin{figure}[H]
\centering
\includegraphics[width=0.8\textwidth]{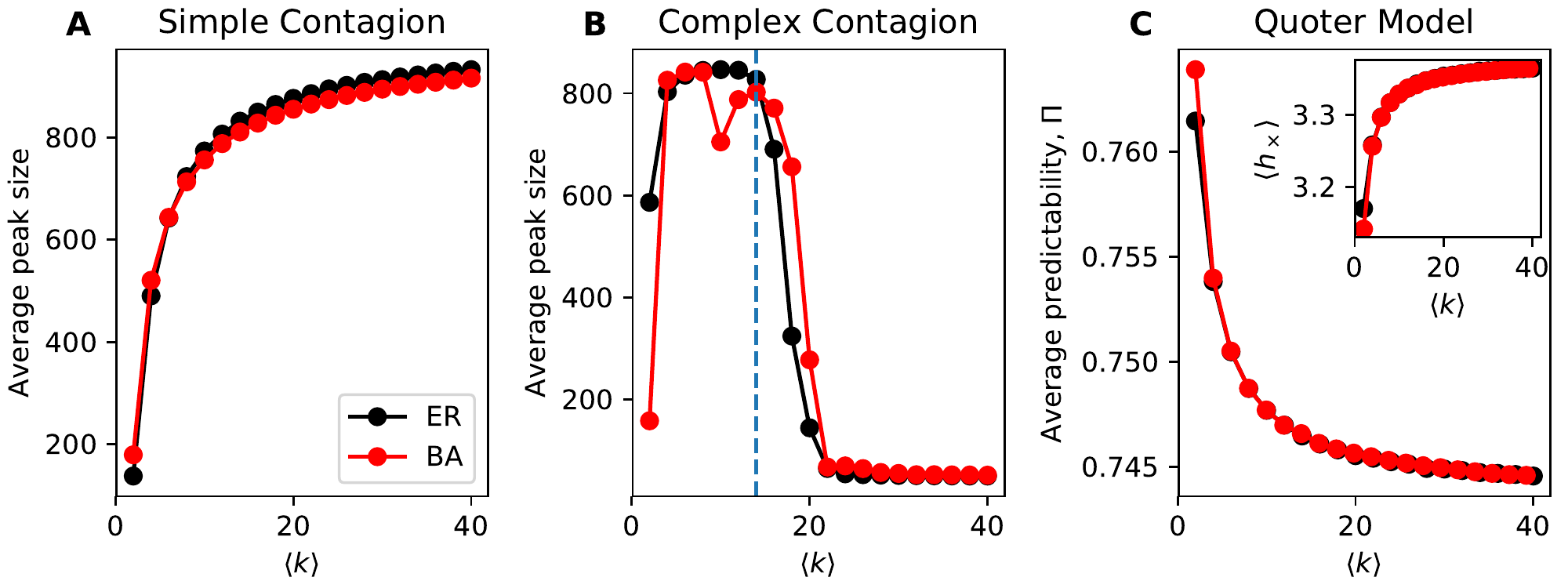}
\caption{%
    Denser networks are associated with higher information flow for simple contagion but lower information flow for both complex contagion and the quoter model. 
    Here density is measured by average degree $\left<k\right>$ for Erd{\H o}s-R{\'e}nyi (ER) \& Barab{\'a}si-Albert (BA) model networks.
     (\textbf{A}) 
     Simple contagion.
     (\textbf{B}) Complex contagion
     (\textbf{C}) Quoter model.
     (Panel C, inset) Average cross-entropy on links; higher cross-entropies correspond to lower predictabilities and lower information flow, unlike for contagions where higher average peak sizes correspond to higher information flow.
     Networks consisted of $N=1000$ nodes and each point constitutes 200 simulations;
     parameters for simulating information flow in these models are described in Section~\ref{sec:methods}.
     \label{fig:densityERBA}}
\end{figure}

\begin{figure}[H]
\centering
\includegraphics[width=0.8\textwidth]{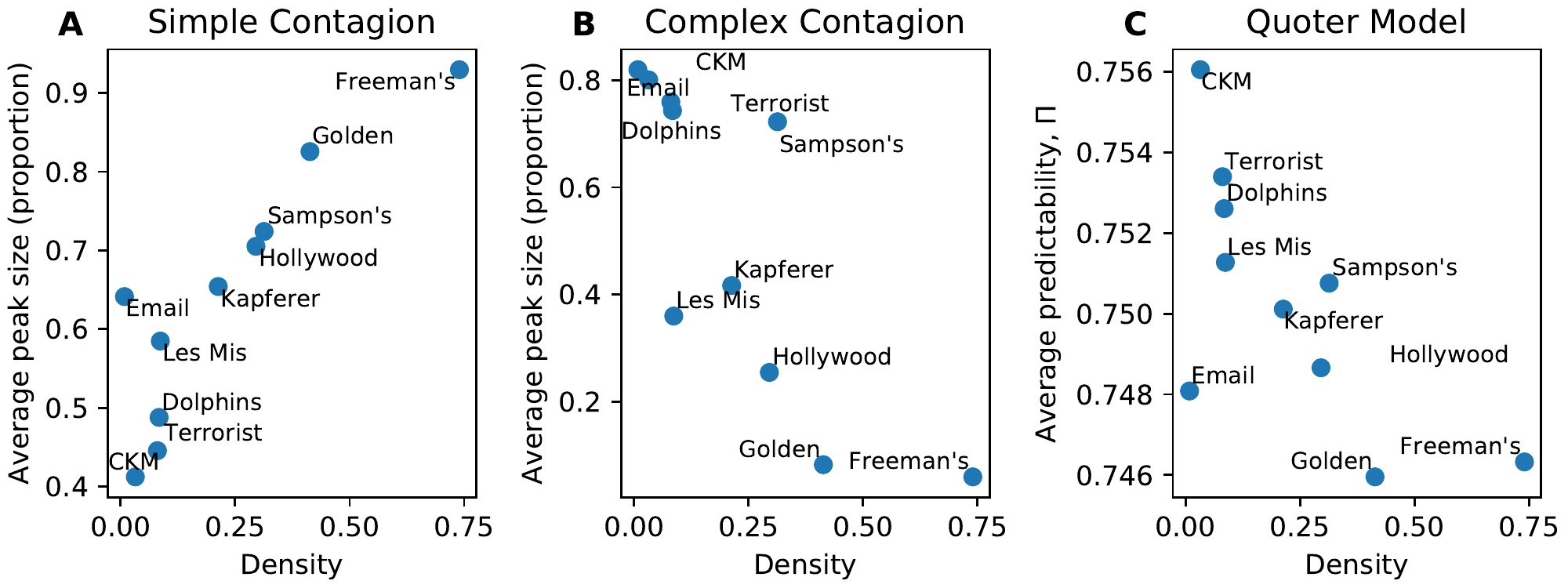}
\caption{%
Information flow  on  real-world networks.
(\textbf{A}) 
     Simple contagion.
     (\textbf{B}) Complex contagion.
     (\textbf{C})
     Quoter model.
     Here information flow measures (average peak size, average text predictability) are compared to network density $M/\binom{N}{2}$.
     The association between information flow and density, either positive (simple contagion) or negative (complex contagion, quoter model), is significant (Wald test on non-zero regression slope, $p < 0.05$).
     Each point constitutes 300 simulations.
     \label{fig:densityRealNetworks}
    }
\end{figure}

A distinguishing feature of simple and complex contagion is that denser networks lead to higher spreading for simple contagion and lower spreading (mostly) for complex contagion.
We illustrate this difference using simulations in Figure~\ref{fig:densityERBA}A,B.
For the simple and complex contagion models we use the average peak size of the outbreak as our measure of information flow in the network, whereas for the quoter model we use the average predictability over links.
The decrease in spreading in complex contagion is due to its social reinforcement mechanism: it is more difficult for a contagion to spread when egos have many alters as more alters must adopt the contagion before the ego does.
Yet we see in Figure~\ref{fig:densityERBA}C that the quoter model, which lacks an explicit social reinforcement mechanism, also exhibits lower information flow at higher density.
Here we measure information flow using predictability on links (Section~\ref{subsec:measuringinfoflowMethods}), which is functionally equivalent (Section~\ref{subsec:measuringinfoflow}) in our simulations to the cross-entropy $h_\times$ (Figure~\ref{fig:densityERBA}C inset).
Please note that while the curve for $h_\times$ looks visually similar to that of simple contagion’s average peak size, it is measuring the opposite effect: higher $h_\times$ corresponds to lower information flow.
These results also hold on our corpus of real-world networks (Figure~\ref{fig:densityRealNetworks}).

\begin{figure}[H]
    \centering
    \includegraphics[width=0.8\textwidth]{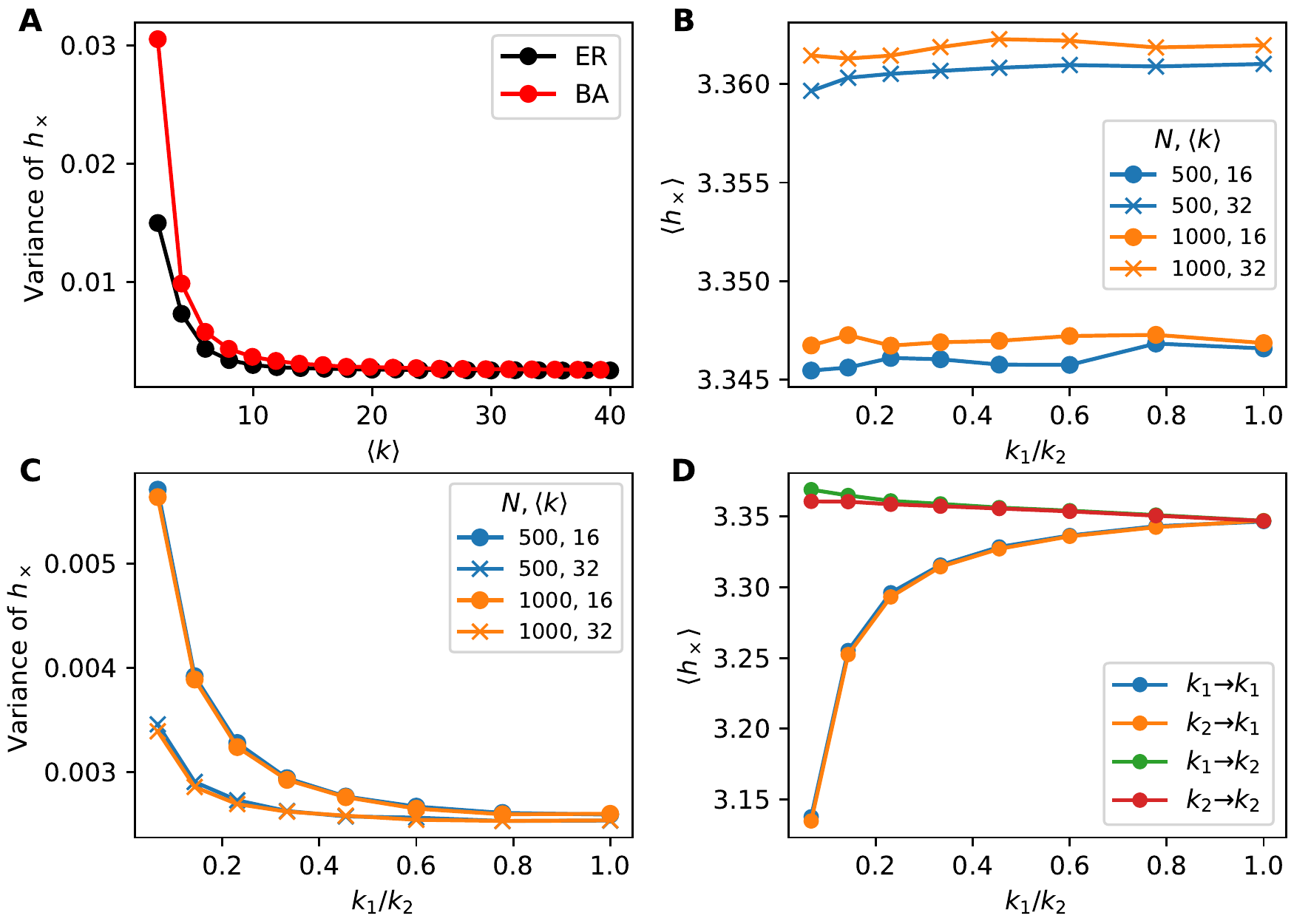}
     \caption{Exploring the variance of information flow.
     (\textbf{A}) Variance of cross-entropy is higher at low densities for BA than ER networks despite the average $h_\times$ being similar (Figure~\ref{fig:densityERBA}C). 
     (\textbf{B--D}) Information flow on dichotomous networks (random networks where all nodes have degree $k_1$ or degree $k_2$, allowing tunable degree heterogeneity) of size $N \in \{500, 1000\}$  with $\langle k \rangle \in \{16, 32\}$. 
     Each point constitutes 500 trials.  
     \textbf{(B)} Average cross-entropy versus $k_1/k_2$. 
     Degree heterogeneity does not affect average cross-entropy, supporting Figure~\ref{fig:densityERBA}C. 
     Network size has a smaller effect on $h_\times$ compared to the average degree.
     \textbf{(C)}
     Variance of cross-entropy versus $k_1/k_2$. 
     Higher degree heterogeneity (lower $k_1/k_2$) leads to higher variation in $h_\times$ over links, indicating the existence of highly predictive nodes and nodes that contribute little predictive information within heterogeneous networks.
     \textbf{(D)} Dichotomous networks of size $N=1000$ and $\langle k \rangle = 16$. Average cross-entropy over links conditioned on degrees of endpoints (predicting ego from alter). 
     Only the degree of the ego matters, approximately, not the degree of the alter.\label{fig:degreeVariance}
     }
\end{figure}

Somewhat surprisingly, in Figure~\ref{fig:densityERBA}C we see that Erd{\H o}s-R{\'e}nyi (ER) and Barab{\'a}si-Albert (BA) networks are qualitatively indistinguishable in terms of information flow, despite the preponderance of hubs in the latter that we expect would play an out-sized role in information flow. 
To better understand this observation, we investigated the variance of $h_\times$ over links in Figure~\ref{fig:degreeVariance}A.
We see that the cross-entropy varies more from link to link in the BA networks than for ER networks, indicating that hubs do not move the average information flow but do create fluctuations in the flow, especially for sparser networks.

To further explore the role of network structure heterogeneity, we investigate dichotomous networks (Section~\ref{subsec:syntheticnetworks}). Here half the nodes have degree $k_1$ and the other half have degree $k_2$. 
Varying the degree ratio $k_1/k_2$ allows us to tune the degree variance within this simplified network model.
In Figure~\ref{fig:degreeVariance}B we see that the total number of nodes and average degree change the average information flow while the degree heterogeneity ($k_1/k_2$) has little effect.
Yet degree heterogeneity does affect the variance of information flow (Figure~\ref{fig:degreeVariance}C).
These simpler dichotomous networks show the same effects as observed previously in BA networks.

The simplified bimodal degree distribution of dichotomous networks also lets us explore the effects of ego and alter degrees by computing conditional expectations of $h_\times$ conditioned on degree.
We see from the grouping of curves in Figure~\ref{fig:degreeVariance}D that the degree of the ego (the node being predicted) but not the alter (the node predicting) plays a role in the information flow: degree-$k_1$ egos have more information flow than degree-$k_2$ egos regardless of the degree of the alter.

\subsection{Interplay of Clustering and Information Flow}
\label{subsec:clustering}

\begin{figure}[H]
    \centering
    \includegraphics[width=0.8\textwidth]{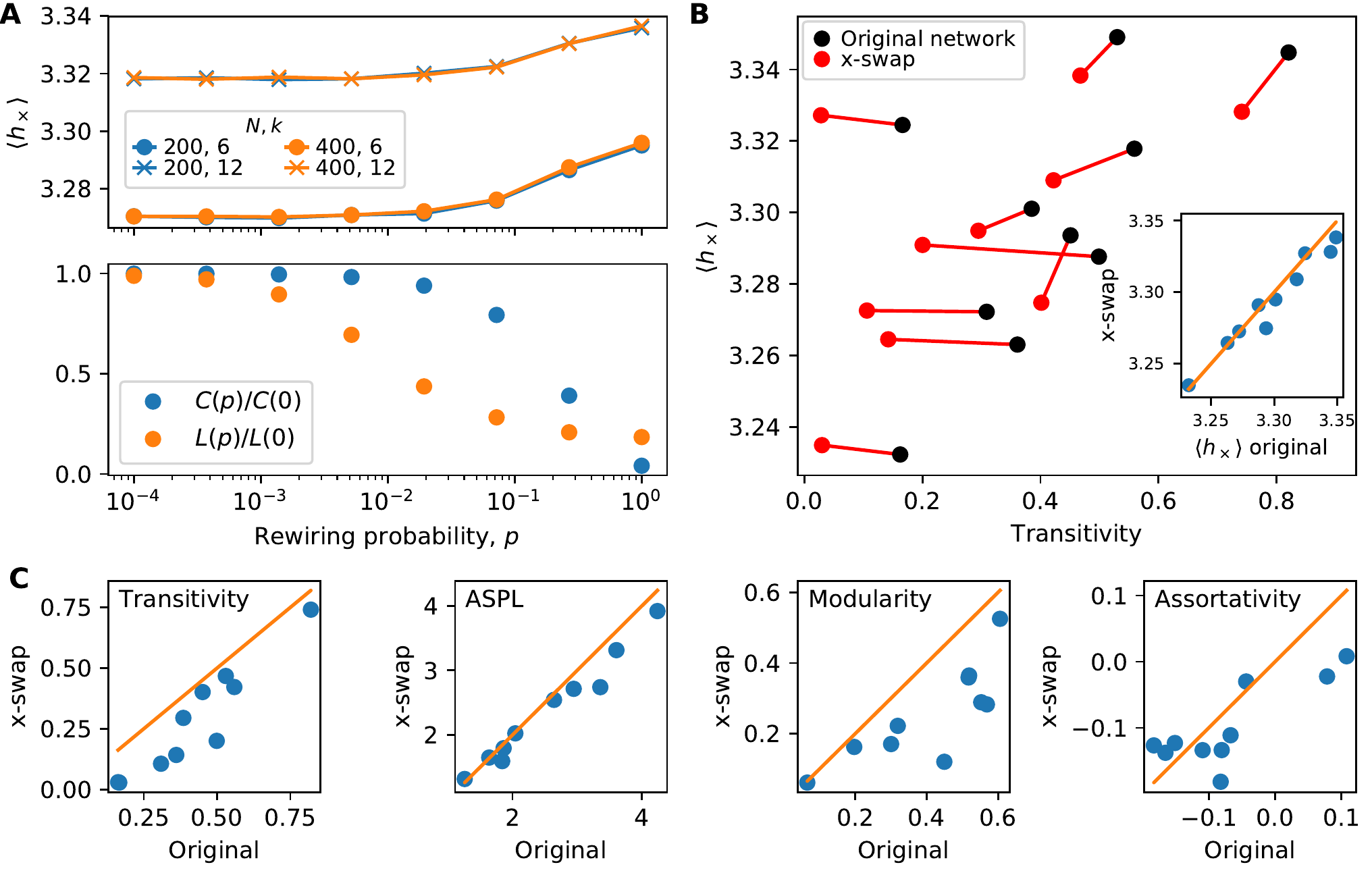}
     \caption{%
     Mixed effects of clustering on information flow.
     \textbf{(A)} Information flow on small-world networks of size $N \in \{200,400\}$ and average degree $k \in \{6,12\}$. 
     As network rewiring increases (and clustering decreases) $h_\times$ increases. 
     This suggests that clustered networks promote information flow.
     Rewiring a small-world network changes the diameter ($L$) as well the clustering (panel A, bottom); however, $h_\times$ begins to increase primarily when the clustering begins to drop, not when diameter begins to drop.
     Each point constitutes 300 trials. 
     \textbf{(B)} 
     Average cross-entropy versus transitivity for real-world networks. 
     By randomizing networks using the standard ``x-swap'' method (Section~\ref{subsec:syntheticnetworks}), we can lower the transitivity and investigate how $h_\times$ changes.
     Some networks show little change in $h_\times$ on randomized networks compared with the original networks, while others show a slight decrease in $h_\times$.
     This is especially visible in the inset comparing $h_\times$ directly. 
     Each point constitutes 300 simulations.
     \textbf{(C)}
    Several network properties before and after the x-swap method. 
    While the x-swap method lowers transitivity, it also alters other important network properties, making it challenging to isolate the role of clustering from other properties.\label{fig:clusteringPromotes} 
    }
\end{figure}

Next, we study how clustering (transitivity) affects information flow.
Clustering plays a complicated role in both simple and complex contagion~\cite{miller2009randomClustered,o2015mathematical} and we report interesting, if mixed, results in Figure~\ref{fig:clusteringPromotes} with the quoter model's information flow.

First, 
in Figure~\ref{fig:clusteringPromotes}A we study information flow for small-world networks that are randomly rewired to remove clustering~\cite{watts1998collective}.
Regardless of network size or average degree, information flow decreases (higher $h_\times$ in top panel of Figure~\ref{fig:clusteringPromotes}A) as clustering decreases (Figure~\ref{fig:clusteringPromotes}A bottom panel).
Please note that rewiring also changes the diameter of the small-world network, but we see that the main increase in $h_\times$ occurs when clustering begins to drop.
In small-world networks, clustering tends to promote information flow.

Next, in Figure~\ref{fig:clusteringPromotes}B we investigate  transitivity in the corpus of real-world networks.
For each network, we compute information flow on the original network and on a replicate of the network that is randomized by the ``x-swap'' method.
The x-swap method lowers transitivity for all networks but for half of the networks it also lowers $h_\times$, contradicting the previous results on small-world networks by indicating that transitivity \textit{inhibits} information.
However, it is challenging to draw a sharp conclusion from this x-swap procedure as it also affects other network properties simultaneously.
We illustrate this in Figure~\ref{fig:clusteringPromotes}C where we compare four network properties in the original and x-swapped networks. 
X-swapping affects transitivity but also average shortest path length (ASPL), modularity and assortativity (degree correlations). 
This means the changes in information flow seen in Figure~\ref{fig:clusteringPromotes}B may be due to changes in a combination of these (and possibly other) network properties.
Unfortunately, it remains an open research problem how best to systematically control for network properties to uncover their effects on dynamics.

\subsection{Community Structure and the Weakness of Long Ties}
\label{subsec:modularity}

The effects of long-range links on information flow have been investigated for some time, from the seminal ``strength of weak ties''~\cite{granovetter1977strength} and the contrasting ``weakness of long ties'' in complex contagion~\cite{centola2007complex}.
Here we investigate long ties in the context of community structure: In networks with densely connected groups of nodes, long ties act to bridge nodes in different groups.
How does information flow differ between groups compared to flow within groups?

\begin{figure}[H]
    \centering
    \includegraphics[width=0.8\textwidth]{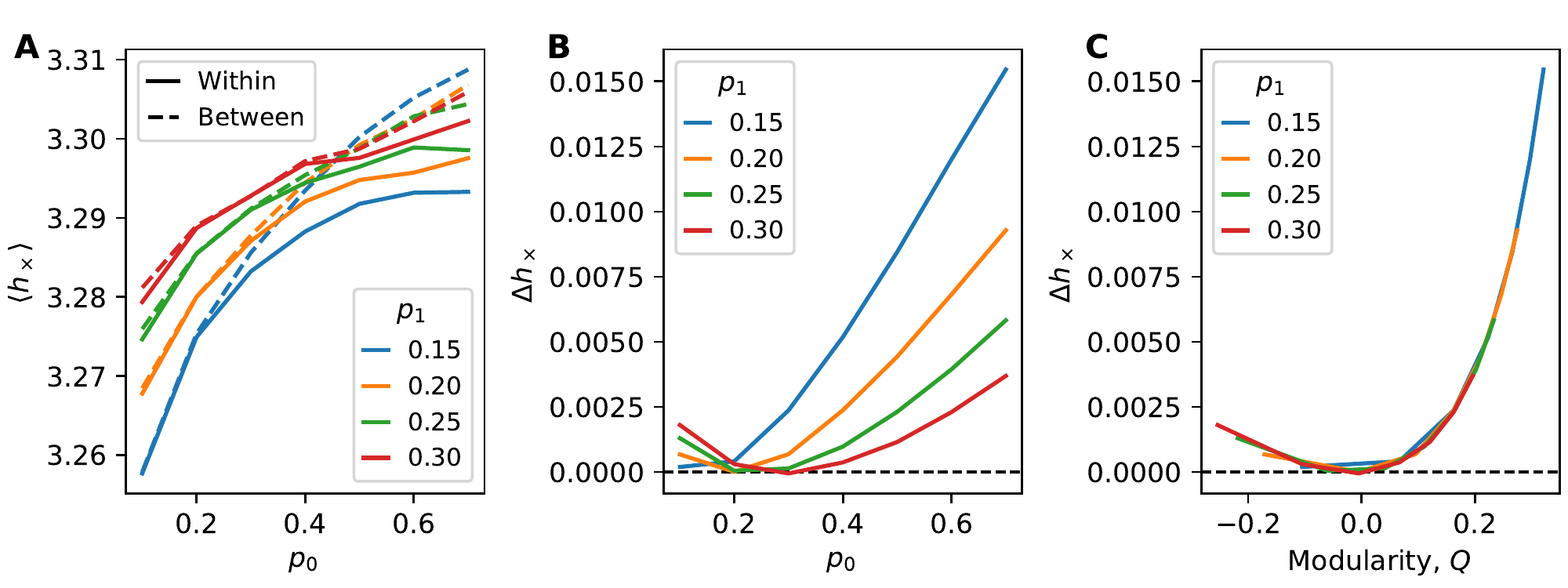}
     \caption{Information flow within the stochastic block model (SBM) of $N=100$ (two blocks of size $N=50$). 
     Each point constitutes 10k trials. 
     \textbf{(A)} 
     Average cross-entropy on within-block edges and between-block edges as a function of the within-block connection probability $p_0$ for different between-block connection probabilities $p_1$.
     (\textbf{B}, \textbf{C}) 
     Examining the cross-entropy difference
    $\Delta h_\times \equiv \langle h_\times(\text{between})\rangle - \langle h_\times(\text{within}) \rangle$ across (\textbf{B}) connection probabilities and (\textbf{C}) modularity $Q$.
    Examining $\Delta h_\times$ as a function of modularity $Q$ shows a clear collapse across values of SBM probabilities. 
    Interestingly, anti-community structure ($Q < 0$) still leads to positive $\Delta h_\times$, indicating that information flow is still more prevalent within blocks.\label{fig:SBM}
    }
\end{figure}

Using the stochastic block model (Section~\ref{subsec:syntheticnetworks}) with two groups of equal size as a model for networks with dense modules, we study in Figure~\ref{fig:SBM} information flow between and within groups.
The two-group SBM is parameterized by two connection probabilities, the probability for a link within each group ($p_0$) and the probability for a link between the two groups ($p_1$). 
In Figure~\ref{fig:SBM}A we see that information flow decreases as $p_0$ increases and the network becomes denser.
Likewise, the difference in information flow $\Delta h_\times$ increases due to between-block links containing less predictive information (Figure~\ref{fig:SBM}B). 
This supports the well-known ``weakness of long ties'' feature of complex contagion.
For larger values of $p_1$, when there are more links connecting the groups making them less distinct, this difference decreases. 
The collapse of curves in Figure~\ref{fig:SBM}C indicates $\Delta h_\times$ is entirely predicated on the network modularity $Q$.

Interestingly, we also remark that $\Delta h_\times$ is always positive---even when $p_0 < p_1$ (equivalently, $Q<0$). 
We would expect more information flow between groups than within when within this ``anti-community'' regime of the SBM, when there are more links between groups than within groups, yet we observe a weak effect otherwise.

\subsection{The Role of Dynamic Heterogeneity}\label{subsec:dynamicheterogeneity}

In our results so far, we have treated nodes as identical within the quoter model and focused only on their topological differences within the network.
Yet recent studies have underlined the importance of comparing dynamic heterogeneity with structural heterogeneity~\cite{deArruda2020impact}.
Here we taken an exploratory step in this direction by considering a generalization of the quoter model where nodes have different vocabulary distributions.

We explored how information flow changes in the stochastic block model when the nodes in the two blocks have different vocabulary distributions.
This is intended to model a difference in the nodes between the two groups, capturing in the quoter model a social homophily in how egos write.
Specifically, we assume they have the same vocabularies and follow Zipf distributions, but the exponent of the Zipf distribution is different: nodes in block A have exponent $\alpha_A$ and nodes in block B have exponent $\alpha_B$. 
A larger $\alpha$ (steeper distribution) corresponds to a less diverse vocabulary, and could capture a group of nodes that is more consistent and repetitive in their dialog.
In contrast, a lower $\alpha$ (shallower distribution) may describe a group of nodes that uses more diverse words.

\begin{figure}[t]
\centering
    \includegraphics[width=0.8\textwidth]{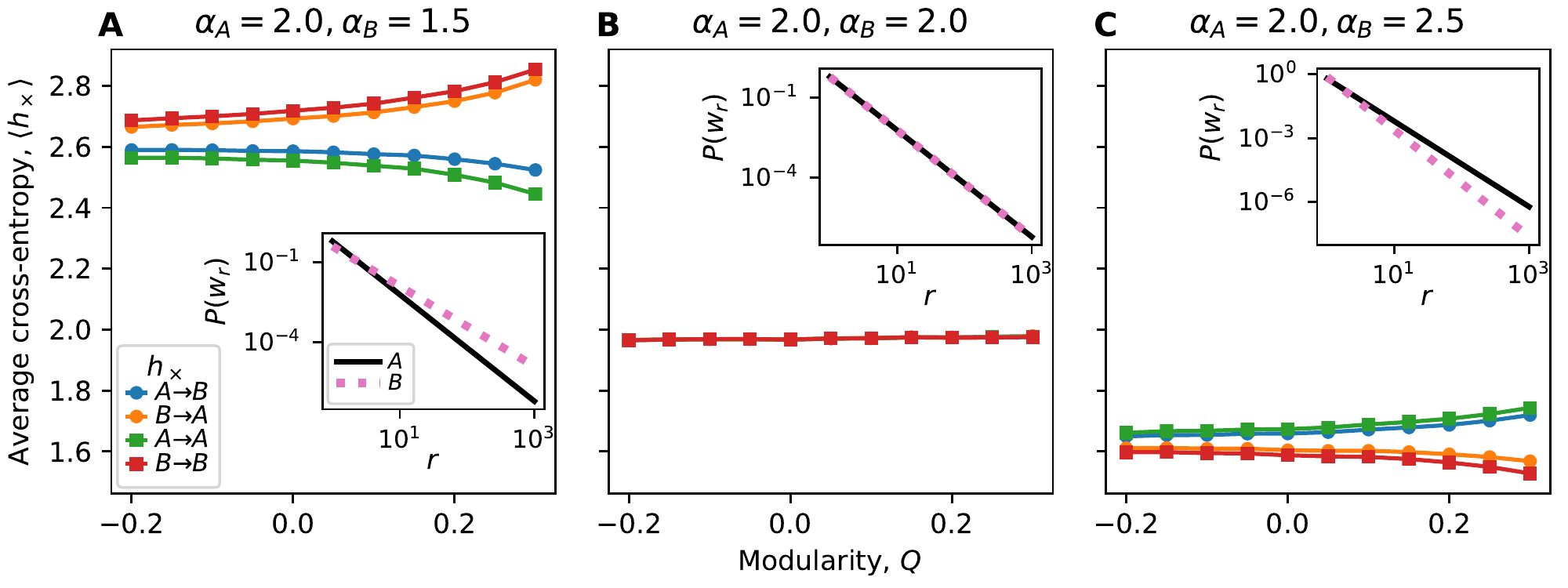}
     \caption{Effects of dynamic heterogeneity on information flow in the stochastic block model. 
     Nodes in group $A$ have Zipfian vocabulary distribution with exponent $\alpha_A$ while nodes in $B$ have exponent $\alpha_B$.
     The between-block connection probability is fixed ($p_1=0.15$) and the within-block connection probability $p_0$ is varied to generate a range of modularities.
     Since the structure is symmetric (subgraphs $A$ and $B$ have the same size and expected density), we only show the result of fixing $\alpha_A=2$ and varying $\alpha_B$.
     Each point constitutes 150 trials.
     \textbf{(A)} The vocabulary distribution of group $A$ has a lower Shannon entropy than of $B$, and this difference is visible from examining links $A\to A$ and $B \to B$. When examining links $A \to B$ and $B \to A$, the cross-entropy is mainly dependent on the vocabulary distribution of the alter. As modularity increases, differences between the predictabilities of various nodes are exaggerated.
     \textbf{(B)} In homogeneous communities, the cross-entropy does not vary with modularity at such a scale.
     \textbf{(C)} The vocabulary distribution of group $A$ has a higher Shannon entropy than of $B$. 
     Similar mirror results are seen as in panel A. 
     \label{fig:vocab_SBM}
     }
\end{figure}

Figure~\ref{fig:vocab_SBM} shows how information flow changes when the two blocks have different vocabulary distributions (Figure~\ref{fig:vocab_SBM}A,C) compared with the same distribution (Figure~\ref{fig:vocab_SBM}B).
For illustration, we show the Zipfian vocabulary distributions for the two groups as insets in Figure~\ref{fig:vocab_SBM}.
We observe a much larger trend in how cross-entropy changes with modularity when the exponents are not equal compared to when they are equal.
This underscores how structural features (the degree of modularity) greatly magnifies the effects of intrinsic dynamic heterogeneity (different vocabulary distributions).
While modularity plays a role even when the two groups have identical vocabulary distributions (Figure~\ref{fig:SBM}), this difference is challenging to detect in Figure~\ref{fig:vocab_SBM}B when viewed on the scale of groups with different vocabulary distributions (Figure~\ref{fig:vocab_SBM}A,C).

\section{Discussion}
\label{sec:discussion}

In this work, we have studied how the social flow of written information can be affected by network properties such as the density of links, preponderance of triangles, and modular or community structure.
We focused on the quoter model, a toy model for a network of individuals to communicate by generating text sequences and applied information-theoretic estimators of the information flow to these texts.
We compared results of information flow in the quoter model with traditional simple and complex contagion models.

A particularly intriguing facet of the interplay between quoter model dynamics and network topology is how the quoter model exhibits both the
density-driven inhibition of information flow
and the weakness of long ties 
that are signatures
of complex contagion, despite lacking an explicit mechanism of social reinforcement.
Social reinforcement, the idea that individuals adopt a piece of information only after receiving repeat exposure from social ties, is considered one of the characteristics that distinguishes complex contagion from epidemic spreading.
Social reinforcement mechanisms better model how people perceive and react to information.
Yet we found here that social reinforcement is not strictly necessary when modeling a more nuanced view of information flow. 
In particular, considering text streams (as generated by the quoter model) and predictive measures of information flow (as quantified using cross-entropy estimators) allows us to capture how information can be ``drowned out'' by the increased ``cross-talk'' that occurs in denser networks, showing how increased density can inhibit information flow.
Further pursuing this line of investigation may give more insight into information flow and even human behavior within social networks.

We also found a mixed combination of results relating clustering to information flow.
For small-world (Watts–Strogatz) networks, increasing the clustering leads to a significant increase in information flow (decrease in cross-entropy).
At the same time, however, experiments on real-world networks showed the opposite effect: randomizing networks to lower transitivity while preserving connectedness and the degree distribution leads to a \textit{decrease} in information flow.
However, this well-established randomization procedure does not control for other network properties such as modularity or average shortest path length, so it remains an open question if the interplay of multiple effects may resolve the discrepancy between these results.

Another interesting result related information flow to community structure, with the modularity $Q$ used to measure the strength of the modular divide.
When $Q > 0$, meaning there were fewer links between modules than expected, we found in Figure~\ref{fig:SBM} an increase in cross-entropy between modules compared with the cross-entropy between nodes that share a module, as expected by the ``weakness of long ties''.
However, we found the same increase in cross-entropy when $Q < 0$, where there were more links between modules than expected. 
We would initially expect this regime of ``anti-community'' structure to have more information flow between modules as there exist more links to facilitate this flow.
One possible reason for this anti-community result is that nodes in the same group, while having fewer direct links to one another, may have many links to common nodes in the other group, leading to more similar inputs to their texts.
This nonlocal interplay of information flow and network structure is an intriguing avenue for future work.

There are some important limitations to discuss regarding this work.
We only considered undirected, unweighted networks. In the context of social networks, this implies all relationships are reciprocal and equal in strength. Future work should extend to directed, weighted networks. 
Furthermore, a more exhaustive study of the robustness of results to parameter choices is necessary (we take a first step towards this in Appendix~\ref{app:robustness}). 
Vocabulary size is another parameter worth exploring; here we assume it is constant across all nodes.
Likewise, cross-entropy (Equation~\eqref{eqn:crossEntropy}) is a somewhat simplistic information-theoretic measure of information flow, and it is important to consider more advanced measures. 
Measures such as transfer or causation entropy can offer more insight, quantifying non-redundant information and allowing us to identify indirect influences~\cite{schreiber2000measuring,sun2014causation}. 
However, in the context of time-ordered social text data, it is challenging to estimate conditional entropies, making it non-obvious how to implement such measures~\cite{quoterModel2018}.
Finally, while we observed several features that are signatures of complex contagion, not all features of complex contagion are exhibited by the quoter model.
For example, there is an optimal modularity that maximizes spreading of complex contagions within the stochastic block model: if $Q$ is either too small or too large then the contagion will not spread~\cite{nematzadeh2014optimal}.
We were unable to observe a corresponding feature within the quoter model.
This warrants further investigation, in particular to understand if this is due to how the quoter model differs from complex contagion models, or if it is due to the information-theoretic measure of information, or a combination of the two.

In general, contagion models are a successful way to study information flow in social networks, but to gain more insight it is necessary to adopt more nuanced views of information flow.
We argue here that information theory can provide a pathway towards these insights, especially when combined with models such as the quoter model that capture features of human behavior while also modeling key aspects of the data being generated by social network platforms. 
\authorcontributions{Conceptualization, T.P., L.M. and J.B.; Funding acquisition, L.M. and J.B.; Investigation, T.P., S.M. and L.M.; Methodology, T.P., T.S. and J.B.; Project administration, J.B.; Software, T.P., T.S. and L.M.; Supervision, L.M. and J.B.; Validation, T.P., S.M. and L.M.; Visualization, T.P.; Writing–original draft, T.P. and J.B.; Writing–review \& editing, T.P. and J.B.. All authors have read and agreed to the published version of the manuscript.}

\funding{This material is based upon work supported by the National Science Foundation under Grant No.\ IIS-1447634.}

\conflictsofinterest{The authors declare no conflict of interest.} 

\abbreviations{The following abbreviations are used in this manuscript:\\

\noindent 
\begin{tabular}{@{}ll}
ASPL & Average Shortest Path Length\\
BA & Barab{\'a}si-Albert\\
ER & Erd{\H o}s-R{\'e}nyi\\
SBM & Stochastic Block Model\\
SI & Susceptible-Infected\\
SIR & Susceptible-Infected-Recovered\\
WS & Watts–Strogatz\\
\end{tabular}}

\appendixtitles{yes} %
\appendix

\section{Further Exploring Quoter Model Parameters}
\label{app:robustness}

\begin{figure}[H]
\centering
    \includegraphics[width=0.8\textwidth]{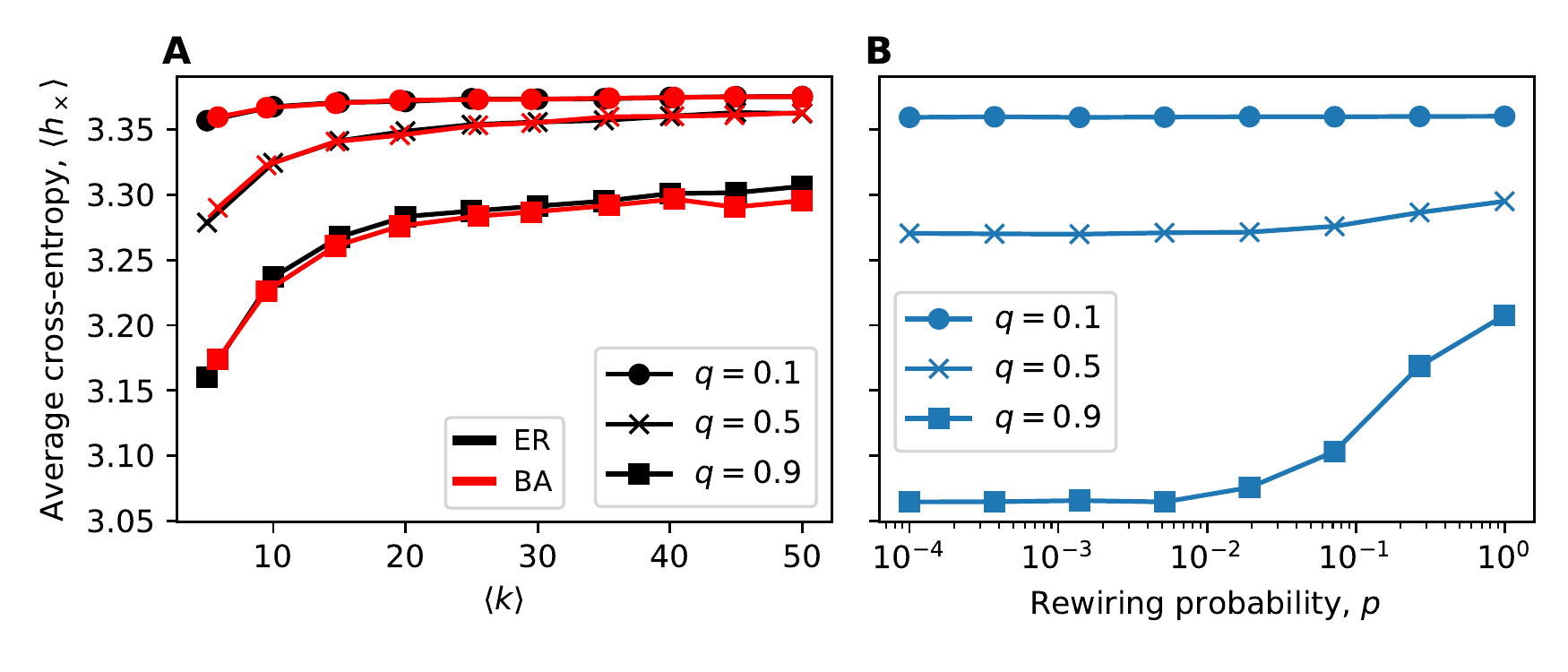}
     \caption{Trends in information flow in ER, BA, and small-world networks for $q\in\{0.1,0.5,0.9\}$.
     Except for very low quote probabilities, we see qualitatively similar trends.
     \textbf{(A)} ER \& BA networks of size $N=100$ with varying average degree.
     Each point constitutes 200 simulations. 
     \textbf{(B)} Small-world networks of size $N=200$ with $k=6$ with varying rewiring probability. 
     Each point constitutes 500 simulations.   
      \label{fig:robustness_ER_BA_q}
    }
\end{figure}

\begin{figure}[H]
    \centering
     \includegraphics[width=\textwidth]{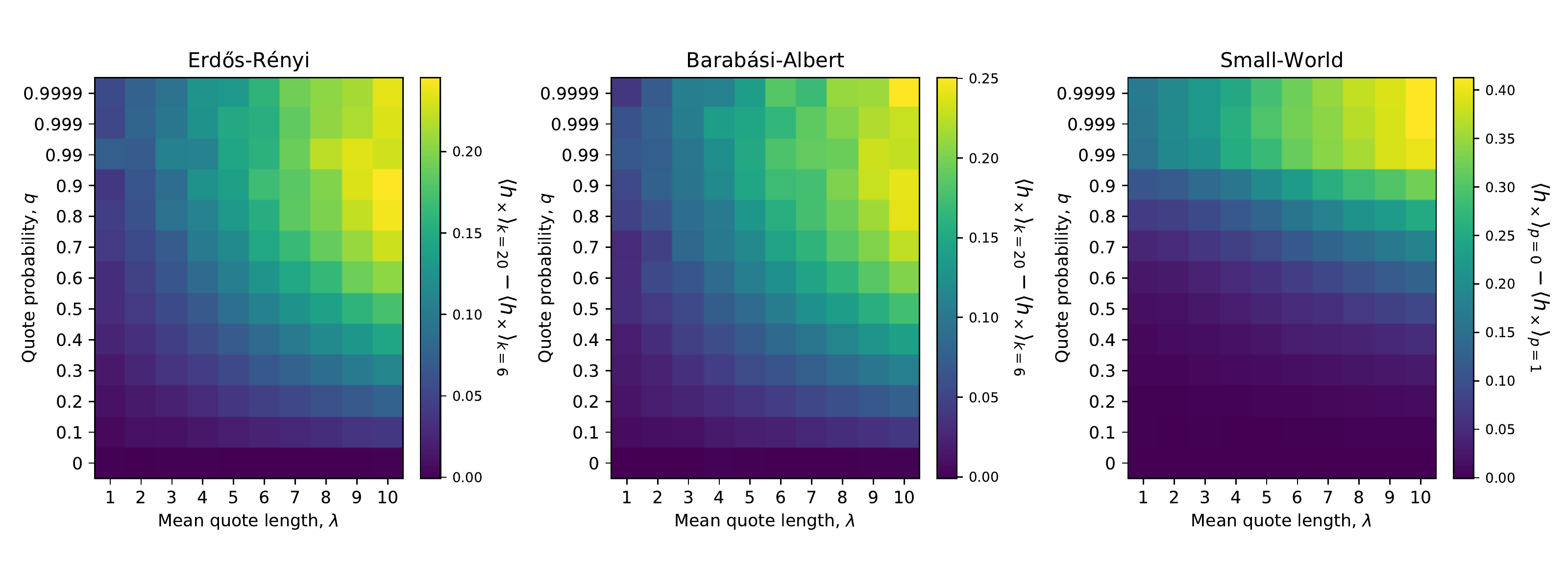}
     \caption{
     Effects of quoter model parameter choices on observed trends.
     Information flow is lower for denser ER and BA networks across a range of $q$ and $\lambda$ with the effect being more pronounced at higher values of $q$ and $\lambda$.
     Likewise, for small-world networks, more clustering (lower $p$) exhibits higher $h_\times$ than less clustering (higher $p$), with the effect being most pronounced at $q > 0.5$ regardless of $\lambda$.
    Here, ER \& BA networks had $N=100$ and small-world networks had $N=200$ and $k=6$. Each cell constitutes 100 simulations.
    \label{fig:robustness_q_lambda}
    }
\end{figure}

To support our results, here we explore other choices of quoter model parameters ($q$ and $\lambda$).  
The simulations are done on smaller networks to make it less computationally expensive to do a wide sweep of the parameter space.
We first simulate the quoter model on ER, BA, and small-world networks for $q \in \{0.1,0.5,0.9\}$ and vary $\langle k \rangle$ or the rewiring probability, $p$, to support results from Section~\ref{subsec:density} and Section~\ref{subsec:clustering}.
We then simulate the ER, BA, and small-world experiments again for various combinations of the quote probability $q$ and mean quote length $\lambda$. 
We evaluate the robustness of results for ER networks as follows. For each combination of $(q,\lambda)$, we calculate the difference $\langle h_\times\rangle_{k=20} - \langle h_\times\rangle_{k=6}$, whereby $\langle h_\times\rangle_{k=20}$ we mean the average cross-entropy on ER networks of average degree $k=20$.
The quantity will be positive if density inhibits information flow. 
This allows us to assess the how the magnitude of our results vary with $(q,\lambda)$, although it does not confirm a monotonic trend holds.
We repeat these calculations with the BA networks and extend them to the small-world networks by replacing $\langle k \rangle$ with $p \in \{0,1\}$. 
In general, we find in Figures~\ref{fig:robustness_ER_BA_q} and \ref{fig:robustness_q_lambda} that our results are qualitatively robust to parameter choices, with the exception of very small values of $q$, as we expect.

\section{Summarizing $h_\times$}
\label{app:distr}
      
\begin{figure}[H]
    \centering
    \includegraphics[width=\textwidth]{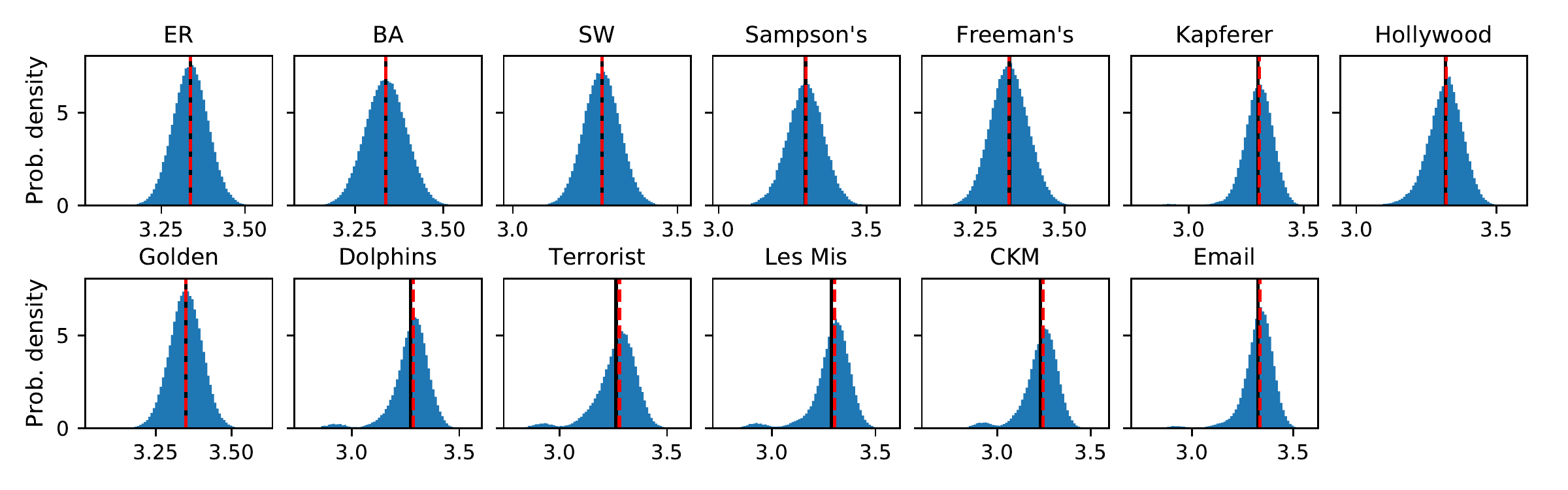}
     \caption{The distributions of $h_\times$ for quoter model simulations on various networks. 
     Examining the distributions supports using $\left<h_\times\right>$ and $\mathrm{Var}(h_\times)$ as summary statistics, although some real networks show a small bimodality (an excess of $h_\times < 3$ bits). 
     We also remark that the mean and median are approximately equal (solid line shows $\langle h_\times \rangle$, dashed line shows median $h_\times$) for all networks.
     ER \& BA networks have $N=1000$ nodes with $\langle k \rangle = 12$, and 200 simulations as in Figure~\ref{fig:densityERBA}. Small-world networks have $N=200$ nodes with $k=6$ and $p=10^{-4}$, and 500 simulations as in Figure~\ref{fig:clusteringPromotes}A. Real-world networks are from 300 simulations as in Figure~\ref{fig:densityRealNetworks} and Figure~\ref{fig:clusteringPromotes}B,C. 
     Quoter model parameters are given in Section~\ref{subsec:simulatingquotermodel}.
     \label{fig:hxDistribution}
     }
\end{figure}

In this work, we summarized $h_\times$ by the mean $\left<h_\times\right>$ and variance $\mathrm{Var}(h_\times)$. 
In Figure~\ref{fig:hxDistribution}, we see that this choice was appropriate: examining the distributions of $h_\times$ for various networks shows that they are approximately normal.
We also find the mean and median $h_\times$ to be approximately equal.

\section{Network Corpus}
\label{app:netcorpus}

All networks studied here can be found through the \href{https://icon.colorado.edu/#!/networks}{Index of Complex Networks} (ICON)~\cite{clauset2016colorado}. We converted any directed or weighted networks to undirected (bidirectional) and unweighted. Details for each of the ten networks:

\begin{enumerate}%
    \item 
    Les Miserables co-appearances \cite{knuth1993stanford} 
    [Undirected, Weighted].
    
    \item 
    Hollywood film music \cite{faulkner1983music} [Undirected, Weighted].
    This is a bipartite network; we converted it to a one-mode projection (nodes are composers and two composers are linked if they worked with the same producer).
    
    \item 
    Freeman’s EIES dataset \cite{freeman1979networkers} [Directed, Weighted].
    We used the ``personal relationships (time 1)'' network.
    
    \item 
    Sampson's monastery \cite{sampson1969novitiate} [Directed, Weighted].
    We used the Pajek dataset. 
    The weight of a directed link represents how an individual rates the other. 
    The rating can be positive (1,2,3 = top 3 ranked) or negative (-1,-2,-3 = worst 3 ranked). 
    We chose to only keep links which were positive.
    
    \item 
    Golden Age of Hollywood \cite{taylor2017eigenvector} [Directed, Weighted].
    We used the aggregated network over 1909-2009.
    
    \item 
    9-11 terrorist network \cite{krebs2002mapping} [Undirected, Unweighted].
    
    \item  
    CKM physicians social network \cite{burt1987social} (1966) [Directed, Unweighted].
    We used ``CKM physicians Freeman'' networks hosted by Linton Freeman, and chose the "friend" network (i.e., the third adjacency matrix). 
    We took only the giant component.
    
    \item Kapferer tailor shop \cite{kapferer1972strategy} (1972) [Undirected, Unweighted].
    We used the ``Kapferer tailor shop 1'' Pajek dataset (kapfts1.dat).
    
    \item Dolphin social network \cite{lusseau2003bottlenose} (1994-2001) [Undirected, Unweighted].
    
    \item Email network (Uni. R-V, Spain, 2003) \cite{guimera2003self} [Directed, Unweighted].
    We used the ``email-uni-rv-spain-arenas'' network. 
\end{enumerate}

\reftitle{References}

%
%

%
%
%

%
%
%
%
%
%
%
%
%
%
%

%
%
%

%
%
%

%
%
%
%
%
%

%
\end{document}